\documentclass[aps,pre,showpacs,preprint]{revtex4}
\usepackage{graphicx}
\usepackage{bm}
\usepackage{amsmath,amssymb}

\begin{document}
\title{Solid-solid transition of the size-polydisperse hard-sphere system}

\author{ Mingcheng Yang}
\author{Hongru Ma}
\email{hrma@sjtu.edu.cn}

\affiliation{Institute of Theoretical Physics, Shanghai Jiao Tong
University, Shanghai 200240, People's Republic of China}
\date{\today}

\begin{abstract}

The solid-solid coexistence of a polydisperse hard sphere system is
studied by using the Monte Carlo simulation. The results show that
for large enough polydispersity the solid-solid coexistence state is
more stable than the single-phase solid. The two coexisting solids
have different composition distributions but the same crystal
structure. Moreover, there is evidence that the solid-solid
transition terminates in a critical point as in the case of the
fluid-fluid transition.

\end{abstract}


\maketitle


The monodisperse hard sphere system is one of the best understood
system in its equilibrium phase behaviors, it undergoes an
entropy-driven first-order transition from a disordered fluid to an
ordered solid as the volume fraction increases \cite{Hoover}. The
system often serves as an excellent starting point to study more
complicated systems and a good model for the description of a class
of colloidal dispersions. However, in a real colloidal system the
particles inevitably exhibit considerable size polydisperseity,
which influences significantly the thermodynamic and dynamic
behaviors of the system \cite{Pusey,Bolhuis,Bartlett,
Fasolo,Phan,Russel,Martin, Schope,Yang1}. Therefore, a more
realistic model describing hard sphere colloids is the
size-polydisperse hard sphere system. The equilibrium phase
behaviors of the polydisperse hard sphere system have not yet been
fully understood. Besides the usual fluid-solid transition, a
general consensus is that there exists a terminal polydispersity,
above which the single-phase crystal becomes thermodynamically
unstable. It is not clear what structure is the thermodynamically
most stable when the polydispersity of the crystal exceeds the
terminal polydispersity. Some theoretical studies show that beyond
the terminal polydispersity the crystal will fractionate into two or
more coexisting solid phases with the same crystal structure
\cite{Bartlett,Fasolo}. Others indicate that a crystal-to-glass
transition will occur \cite{Chaudhuri}. Experimentally, the
crystallization does not occur at large enough polydispersity
\cite{Pusey}. From the experiment we draw no definite conclusion
about the equilibrium phase behavior because the nonequilibrium
effect will dominate the system for such a high polydispersity. So
far, the solid-solid coexistence of polydisperse hard sphere system
is only a theoretical prediction. In order to confirm the
prediction, carefully designed  experimental studies and
comprehensive computer simulations are needed. To the best of our
knowledge, the only simulation evidence up to now comes from the
work by Fern\'{a}ndez et al. \cite{Fernandez}. In that work they
investigated the phase equilibria of the polydisperse soft-sphere
system, and found that the crystal is highly inhomogeneous at large
polydispersity. However, this does not provide us any details about
the inhomogeneous structure.

In this letter, we use the Monte Carlo method to investigate the
solid-solid transition of a polydisperse hard sphere crystal. To
simulate a polydisperse crystal we employ the semigrand ensemble,
which is the best frame to study the polydisperse crystal
\cite{Bolhuis,Kofke,Bates}. In the ensemble the composition
distribution is \emph{a priori} unknown, and the independent
variables are the chemical potential differences $\Delta\mu(\sigma)$
of particles of each kind to a reference kind, which are given in
advance. As a result, in the ensemble the conditions for coexistence
of two phases will be satisfied if only the two phases have the same
pressure and referenced chemical potential. As is well known, the
first-order transition of a hard sphere system can be identified by
looking for a double-peak structure in the volume histogram. Then,
by tuning the pressure one can easily detect the transition point
where the two peaks have equal weight \cite{Borgs}. In actual
computations we locate the transition point using a more tractable
criteria ``equal peak height'' \cite{Challa}, which differs from
equal weight in the finite-size effect, and  gives the same result
in the thermodynamic limit. In the following we describe in detail
the simulation method and discuss the obtained results.

The quantity $P_{iso}(V)=\Upsilon(V)\texttt{exp}(-\beta PV)$
measures the probability density to find a system with the volume
$V$ and the pressure $P$, here $\Upsilon(V)$ is the semigrand
canonical partition function with given differences of excess
chemical potential, defined as
\begin{eqnarray}
\Upsilon(V)
=\frac{1}{N!\Lambda^{3N}(\sigma_{r})}\int_{\sigma_{1}}\cdots\int_{\sigma_{N}}
Z_{N}(V)\times
\exp\left\{\beta\sum_{i=1}^{N}(\mu_{ex}(\sigma_{i})-\mu_{ex}(\sigma_{r}))\right\}
\prod_{i=1}^{N}d\sigma_{i},
\end{eqnarray}
here, $\sigma_{i}$ and $\sigma_{r}$ are the diameter of the $i$th
particle and an arbitrarily chosen referenced component,
respectively. $Z_{N}$ is the canonical configuration integral
$Z_{N}(V)=\int_{V}\cdots\int_{V}e^{-\beta
U}\prod_{i=1}^{N}d\textbf{r}_{i} $, and
$\mu_{ex}(\sigma_i)=\mu(\sigma_i)-kT\ln(\frac{N\Lambda(\sigma_i)^{3}}{V})$
is the excess chemical potential relative to the ideal gas. We now
consider the coexistence of two solids having the same structure, in
this case the two solids can be connected with a reversible path as
in the case of fluid-fluid coexistence. The path can be constructed
in such a way by introducing an extended semigrand canonical
ensemble with the volume as the ensemble variable. By changing the
volume  the system transforms from one coexisting solid to the
other, provided that the solid-solid transition does exist. When we
perform a random walk in the volume space, the probability density
of finding the system in volume $V$ is given by
\begin{equation}
P_\texttt{ext}(V)\propto  {\Upsilon(V,N,T,\Delta\mu_{ex}(\sigma))}.
\end{equation}
The semigrand canonical partition function $\Upsilon(V)$ can be
simulated by using the flat histogram methods
\cite{Berg,F.Wang,S.Wang} to within an overall multiplier, which is
not needed in determine the relative probability. The simulation
involves three kind of moves: particle displacement, particle
resizing and volume changing.

Previous calculations \cite{Yang} show that face-centered-cubic(fcc)
phase is still the most stable for the hard sphere crystal with a
low size-polydispersity. Therefore, both coexisting solids are
considered to be the fcc structure. At present, we still have no
definite knowledge about the position of polydispersity leading to
instability of the single-phase crystal. Recent studies on the
elasticity of the fcc polydisperse hard sphere crystal found that
there exists a mechanical terminal polydispersity(MTP) above which
the crystal is mechanically unstable \cite{Yang1}. The MTP is an
upper limit of the thermodynamical terminal polydispersity, thus we
expect that a single-phase crystal is in the thermodynamically
metastable state at polydispersity slightly below the MTP. The
chemical potential difference function of the metastable crystal can
be calculated with the SNERP algorithm \cite{Yang,Wilding1}. And the
solved chemical potential difference function is then used to study
the solid-solid phase transition. To determine the MTP from elastic
constants, a large amount of simulation time is needed. Here, we
adopt a simple but effective approach, by which the MTP can be
estimated roughly. The approach is based on the observation that,
once the polydispersity is higher than the MTP, the SNERP procedure
will converge very slowly or even does not converge. As a result,
the MTP can be determined roughly by monitoring the chemical
potential difference calculation with SNERP algorithm. Once we know
the chemical potential difference function of a metastable crystal,
we can use it to calculate the probability densities
$P_\texttt{ext}(V)$ and $P_\texttt{iso}(V)$.

\begin{figure}
\includegraphics[angle=0,width=0.42\textwidth]{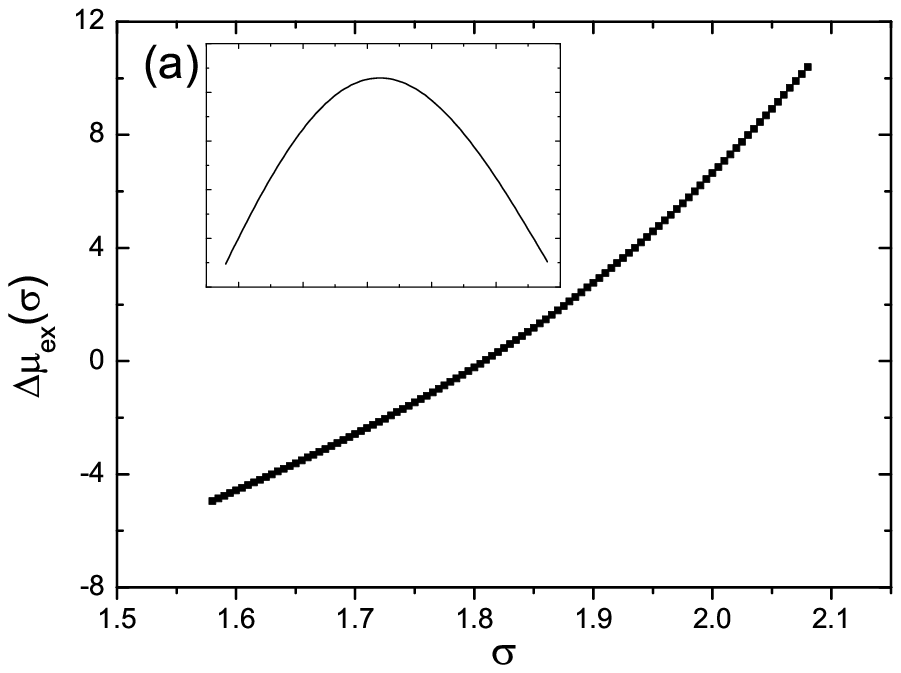}
\includegraphics[angle=0,width=0.42\textwidth]{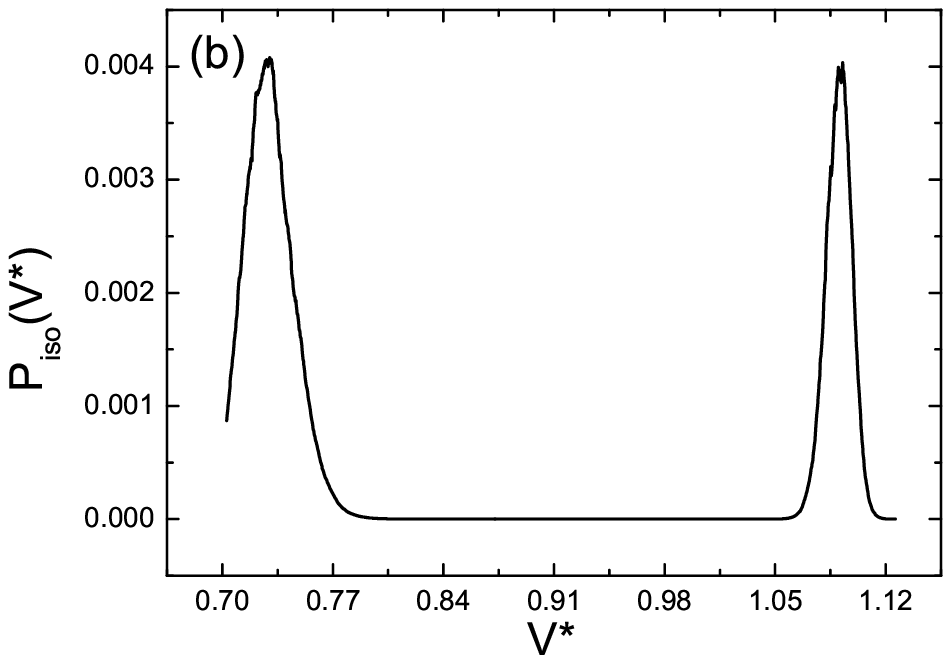}
\caption{(a) the chemical potential difference of an initial
metastable crystal as function of the diameter of particles. It is
obtained from the SNEPR method. The inset shows the prescribed
composition distribution of the metastable crystal. (b) the
distribution of the dimensionless volume of the system at the
coexisting pressure. It is calculated from the extended ensemble
simulation by using the $\Delta\mu_{ex}(\sigma)$ plotted in (a).}
\label{fig1}
\end{figure}

\begin{figure}
\includegraphics[angle=0,width=0.42\textwidth]{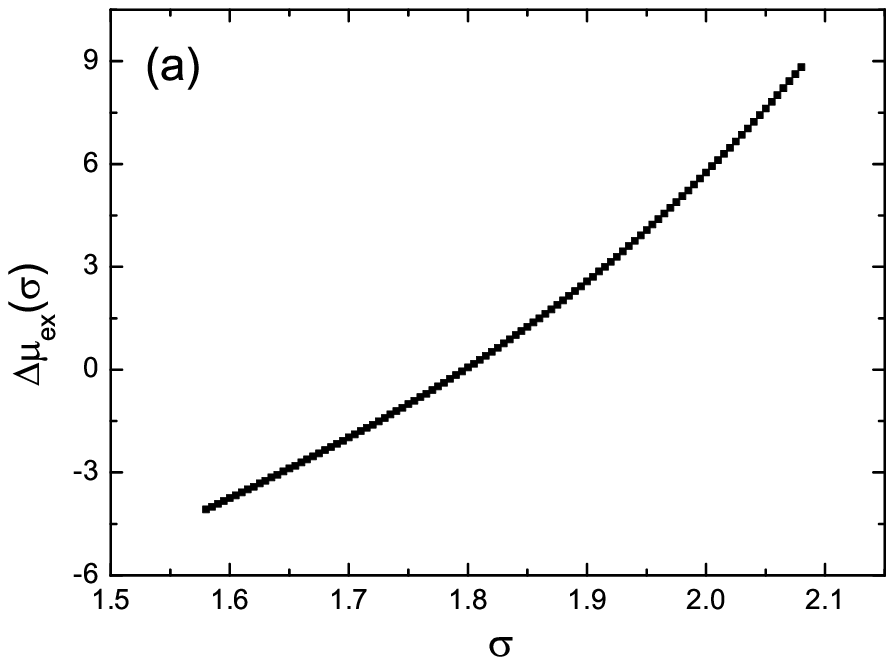}
\includegraphics[angle=0,width=0.42\textwidth]{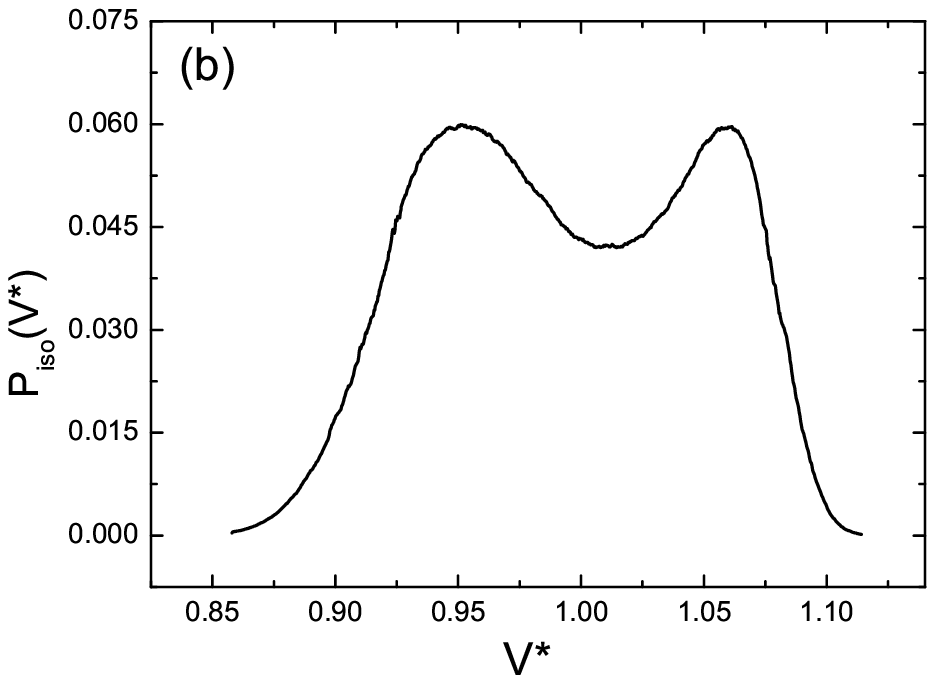}
\caption{The same as Fig. \ref{fig1}, the volume fraction of the
initial metastable crystal is less than that in Fig. \ref{fig1}.
}\label{fig2}
\end{figure}

\begin{figure}
\includegraphics[angle=0,width=0.42\textwidth]{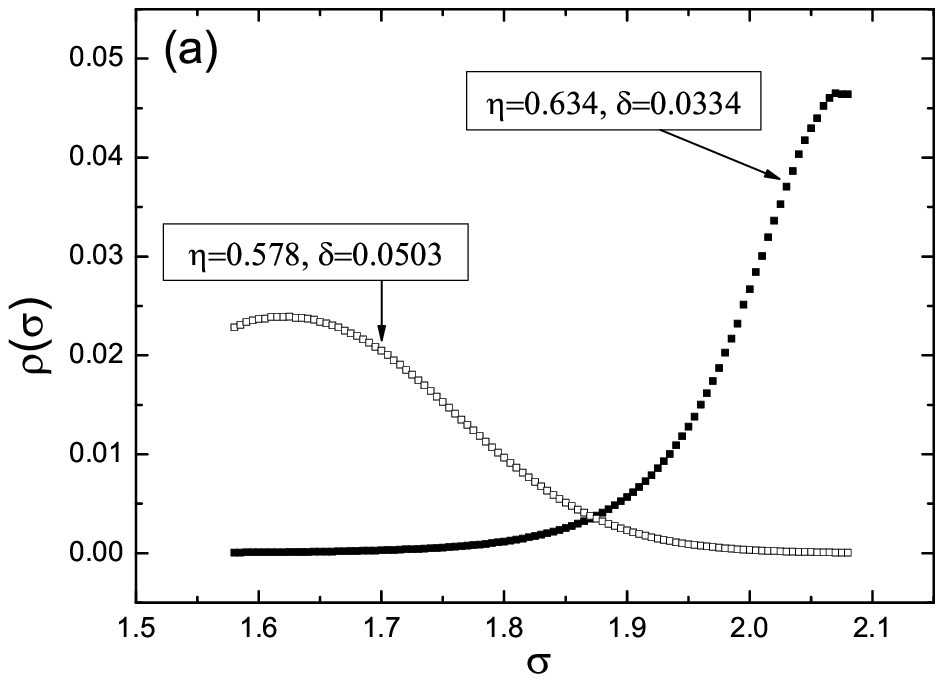}
\includegraphics[angle=0,width=0.42\textwidth]{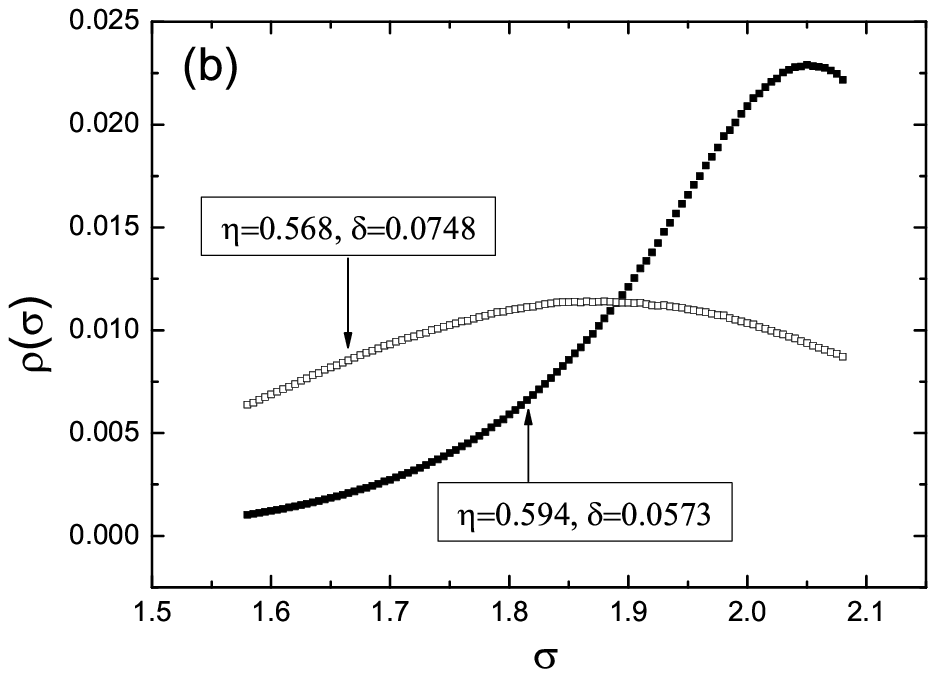}
\caption{(a) the composition distribution of the coexisting solids
given in Fig. \ref{fig1}(b). The solids with the lower(open square)
and higher(filled square) volume fractions correspond to the left
and right peaks in Fig. \ref{fig1}(b), respectively. This is because
the large volume accommodates easily the larger particles. Here,
$\eta$ is the volume fraction and $\delta$ is the polydispersity.
(b) the composition distribution of the coexisting solids given in
Fig. \ref{fig2}(b).}\label{fig3}
\end{figure}

\begin{figure}
\includegraphics[angle=0,width=0.5\textwidth]{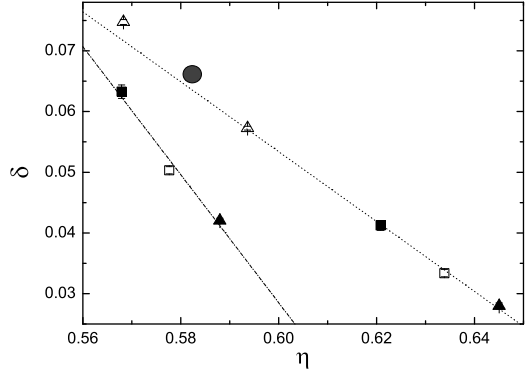}
\caption{The solid-solid coexistence plotted in $\eta-\delta$ plane.
Two coexisting phases are denoted by the same symbols. Dashed line
is a linear fit for the coexisting solids with the lower volume
fraction(except the uppermost triangle); dotted line is a linear fit
for the coexisting solids with the higher volume fraction; filled
circle gives a guess for the critical point.}\label{fig4}
\end{figure}

We consider four initial metastable crystals at the same value for
the composition distribution, but with different volume fractions.
Correspondingly, we have four different chemical potential
difference functions. The initial prescribed composition is a
truncated Schultz function, and the initial configuration is an
ideal fcc crystal. All the simulations are performed with a system
of $256$ size-polydisperse hard spheres in a parallelepiped box,
periodic boundary conditions are used in all three directions.
Figure \ref{fig1}(a) and figure \ref{fig2}(a) are two typical
chemical potential difference functions of the initial metastable
crystals. They are obtained by applying the SNEPR method to the
initial crystals. The initial crystal in Fig. \ref{fig1} has higher
volume fraction than the one in Fig. \ref{fig2}, so its chemical
potential difference is larger. Figure \ref{fig1}(b) and figure
\ref{fig2}(b) show the distributions of the dimensionless volume of
the system, which are calculated from the extended semigrand
canonical ensemble simulation by using the solved chemical potential
difference functions plotted in Fig. \ref{fig1}(a) and Fig.
\ref{fig2}(a), respectively. The dimensionless volume is defined as
$V^{*}=V/N(\overline{\sigma})^3$,
here $\overline{\sigma}$ is the average of the diameter of particles
of the initial metastable crystal. The distributions clearly exhibit
a double-peak structure for a range of pressure, which is the sign
of the existence of solid-solid coexistence. At the coexisting
pressure the two peaks have equal height, each peak represents a
coexisting solid, and the volumes of two coexisting solids are
determined from the positions of the peak maximum. The composition
distribution of each coexisting solid can be obtained from an
additional semigrand canonical simulation by the same chemical
potential difference function. Figure \ref{fig3} displays the
particle size  distributions of the coexisting solids. The
compositions of two coexisting solids are significantly different,
i.e. the fractionation effect is sufficiently large. The coexisting
solid with lower volume fraction has a larger polydispersity than
the one with higher volume fraction, which is consistent with the
results by Fasolo et al \cite{Fasolo}. In the simulation we find
that the separation between the two peaks decreases as the volume
fraction of initial solid decreases, as shown in Fig. \ref{fig1}(b)
and Fig. \ref{fig2}(b). We expect that the two peaks will completely
merge together under some special conditions. This gives the
solid-solid critical point, sketched by the filled circle in Fig.
\ref{fig4}. The early studies also indicated that a monodisperse
system of hard spheres with a short-ranged attractive interaction
undergoes a solid-solid transition with the same crystal structure,
and has a solid-solid critical point \cite{Bolhuis2,Tejero}. We
argue that the critical phenomena of the polydisperse crystal is not
experimentally observable, since in a real colloidal crystal the
particles are not permitted to exchange their positions and to
change their sizes.

Figure \ref{fig4} shows the coexisting solids plotted in the
polydispersity and volume fraction plane. Even though we do not
provide the cloud points(The estimate of cloud points need more
simulation time and techniques \cite{Buzzacchi}.), from the figure
we can draw some important conclusions. For the systems under
consideration we note that, except the lower density solid in Fig.
\ref{fig3}(b)(the uppermost triangle in Fig. \ref{fig4}), the
coexisting phases with lower volume fraction have similar size
distribution function and follow a linear relation. So the linear
fit roughly resembles a cloud curve(dashed line). On the right side
of the cloud curve the single-phase solid becomes thermodynamically
unstable. The cloud curve has a negative slope, so for each volume
fraction of interest there exists a maximum polydispersity, above
which the solid-solid coexistence occurs. On the other hand, all the
coexisting solids with higher volume fraction also have the similar
sizes distribution. Thus we get a second approximate cloud curve by
a linear fitting(dotted line). Comparing with the first case, the
unstable region lies on the left side of the cloud curve, and for
each volume fraction under consideration there exists a minimum
polydispersity stabilizing a single-phase solid. Because the
monodisperse crystal is stable in the volume fraction range of the
simulation, the cloud curve will bend back at the lower
polydispersity, where the fluid-solid transition occurs. However,
with further increasing the polydispersity the cloud curve will bend
back toward the high volume fraction due to the existence of the
terminal polydispersity. The cloud curve looks like a ``reverse
-Z''.

In summary, we investigated the solid-solid transition of the size
polydisperse hard sphere system by the Monte Carlo simulation. The
results indicate that at sufficient high polydispersity the
single-phase solid become unstable and the solid-solid coexistence
will occur. Two coexisting solid phases have different composition
distributions. We also find that the solid-solid critical point can
exist under some special conditions. Though we do not give the cloud
points, we can obtain some qualitative understanding about the
solid-solid phase diagram. It is possible to calculate the cloud
points in the current frame \cite{Buzzacchi}. The existence of the
solid-solid coexistence state can not preclude the occurrence of the
glass transition, but we prefer to think of the glass transition as
a kinetic phenomenon.

The work is supported by the National Natural Science Foundation of
China under grant  No.10334020  and in part by the National Minister
of Education Program for Changjiang Scholars and Innovative Research
Team in University.


\begin{thebibliography}{99}

\bibitem{Hoover}W.G. Hoover and F. H. Ree, J. Chem. Phys. \textbf{49}, 3609 (1968).

\bibitem{Pusey}P. N. Pusey and W. van Megen, Nature (London) \textbf{320}, 340 (1986).

\bibitem{Bolhuis}P.G. Bolhuis and D. A. Kofke, Phys. Rev. E \textbf{54}, 634 (1996);
D. A. Kofke and P.G. Bolhuis, \emph{ibid}. \textbf{59}, 618 (1999).

\bibitem{Bartlett}P. Bartlett, J. Chem. Phys. \textbf{109}, 10970 (1998).

\bibitem{Fasolo}M. Fasolo and P. Sollich, Phys. Rev. Lett. \textbf{91}, 068301 (2003).

\bibitem{Phan}S. Phan, W. B. Russel, Z. Cheng, J. Zhu, P. M. Chaikin, J. H. Dunsmuir,
and R. H. Ottewill, Phys. Rev. E \textbf{54}, 6633 (1996).

\bibitem{Russel}S. Phan, W. B. Russel, J. Zhu and P. M. Chaikin, J. Chem. Phys. \textbf{108}, 9789 (1998).

\bibitem{Martin}S. Martin, G. Bryant, and W. van Megen, Phys. Rev. E \textbf{67}, 061405 (2003).

\bibitem{Schope}H. J. Schope, G. Bryant and W. van Megen, Phys. Rev. Lett. \textbf{96}, 175701 (2006).

\bibitem{Yang1}M. C. Yang and H. R. Ma,  Phys. Rev. E (in press).

\bibitem{Chaudhuri}P. Chaudhuri, S. Karmakar, C. Dasgupta, H. R. Krishnamurthy,
and A. K. Sood, Phys. Rev. Lett. \textbf{95}, 248301 (2005).

\bibitem{Fernandez}L. A. Fern\'{a}ndez, V. Mart\'{i}n-Mayor and P. Verrocchio, Phys.
Rev. Lett. \textbf{98}, 085702 (2007)

\bibitem{Kofke}D. A. Kofke and E. D. Glandt, J. Chem. Phys. \textbf{87}, 4881 (1987).

\bibitem{Bates}M. A. Bates and D. Frenkel, J. Chem. Phys. \textbf{109}, 6193 (1998).

\bibitem{Borgs}C. Borgs and R. Koteck\'{y}, J. Stat. Phys. \textbf{61}, 79 (1990); Phys.
Rev. Lett. \textbf{68}, 1734 (1992).

\bibitem{Challa}M. S. S. Challa, D. P. Landau and K. Binder, Phys. Rev. B. \textbf{34}, 1841 (1986).

\bibitem{Yang}M. C. Yang and H. R. Ma,  J. Chem. Phys. \textbf{128}, 134510 (2008).

\bibitem{Wilding1}N. B. Wilding, J. Chem. Phys. \textbf{119}, 12163 (2003).

\bibitem{Smit}D. Frenkel and B. Smit, \emph{Understanding Molecular
Simulation} (Academic, San Diego,1996).

\bibitem{Briano}J. G. Briano and E. D. Glandt, J. Chem. Phys. \textbf{80}, 3336 (1984).

\bibitem{Lyubartsev}A. P. Lyubartsev, A. A. Martsinovski, S. V. Shevkunov, and P. N.
Vorontsov-Velyaminov, J. Chem. Phys. \textbf{96}, 1776 (1992).

\bibitem{Berg}B. A. Berg and T. Neuhaus, Phys. Rev. Lett. \textbf{68}, 9 (1992).

\bibitem{F.Wang}F. Wang, D. P. Landau, Phys. Rev. Lett. \textbf{86}, 2050 (2001);
Phys. Rev. E \textbf{64}, 056101 (2001).

\bibitem{S.Wang}J. S. Wang and R. H. Swendsen, J. Stat. Phys. \textbf{106}, 245
(2002).

\bibitem{Bolhuis2}P.G. Bolhuis and D. Frenkel, Phys. Rev. Lett. \textbf{72}, 2211 (1994).

\bibitem{Tejero}C. F. Tejero, A. Daanoun, H. N. W. Lekkerkerker, and M. Baus,
Phys. Rev. Lett. \textbf{73}, 752 (1994).

\bibitem{Buzzacchi}M. Buzzacchi, P. Sollich, N. B. Wilding and M.
M\"{u}ller, Phys. Rev. E. \textbf{73}, 046110 (2006).

\end{thebibliography}
\end{document}